\documentclass{article}
\usepackage[left=1.625in,right=1.625in,top=1in,bottom=1.625in]{geometry}
\usepackage[utf8]{inputenc}
\usepackage{graphicx}   
\usepackage{xcolor}     	
\usepackage{caption}    
\usepackage{cite}
\usepackage{lineno}      
\usepackage{amsmath}
\usepackage{soul}        
\usepackage{authblk}
\usepackage{hyperref}
\usepackage{etoolbox}
\newcommand*{\email}[1]{%
    \normalsize\href{mailto:#1}{#1}\par
    }
    
\title{Integrated O- and C-band Silicon-Lithium Niobate Mach-Zehnder Modulators with 100 GHz Bandwidth, Low Voltage and Low Loss }
\author{Forrest Valdez$^{1,*}$, Viphretuo Mere$^{1}$, Xiaoxi Wang$^{1}$, and Shayan Mookherjea$^{**}$}
\affil[1]{University of California, San Diego, Department of Electrical and Computer Engineering, 9500 Gilman Drive, MC 0407, La Jolla, California, USA}
\affil[*]{\email{fgvaldez@eng.ucsd.edu, $^{**}$smookherjea@ucsd.edu}}
 \date{}
 
\begin{document}

\maketitle


\begin{abstract}
Broadband integrated thin-film lithium niobate (TFLN) electro-optic modulators (EOM) are desirable for optical communications and signal processing in both the O-band (1310 nm) and C-band (1550 nm). To address these needs, we design and demonstrate Mach-Zehnder (MZ) EOM devices in a hybrid platform based on TFLN bonded to foundry-fabricated silicon photonic waveguides. Using a single silicon lithography step and a single bonding step, we realize MZ EOM devices which cover both wavelength ranges on the same chip. The EOM devices achieve 100 GHz EO bandwidth (referenced to 1 GHz) and about 2-3 V.cm figure-of-merit ($V_\pi L$) with low on-chip optical loss in both the O-band and C-band. 
\end{abstract}

\section{Introduction}
There is considerable interest in designing and manufacturing thin-film lithium niobate (TFLN) Mach-Zehnder (MZ) electro-optic modulator (EOM) devices, based on the simultaneous achievements of high optical confinement, high EO bandwidth, low voltage, and low optical loss that have been demonstrated recently \cite{sohler2008integrated,Boes2018,wang2018integrated,he2019high,zhu2021integrated}. These attributes are critically important for both optical communications and optical signal processing, and for the development of multi-functional integrated photonic circuits with capabilities beyond a traditional silicon (Si) photonics platform. \textcolor{black}{While the majority of TFLN based modulators have been designed for the low-loss C-band (around 1550 nm) for long-haul communication applications, the O-band (around 1310 nm) is important for short-range optical fiber communications and has lower dispersion \cite{IEEE2015}.} When designing devices for different wavelengths on a common platform, it is challenging to achieve precise RF-optical index matching, since sub-micron-scale waveguides in high-index contrast platforms  are highly dispersive (i.e., the effective modal index, $n_\textrm{eff}$, and the modal cross-sectional area, $A_\textrm{eff}$, vary with wavelength) compared to diffused LN waveguides. There have already been some notable efforts to design TFLN MZ-EOM devices using a common approach at both these bands, e.g., by Stenger et al. \cite{stenger2013engineered} and Sun et al. \cite{sun2020hybrid}; however, state-of-the-art performance has not yet been demonstrated in this way. A unified O-band and C-band 100-GHz-class, few-volt integrated EOM device platform may not only benefit wideband optical communications, but also help in optical signal processing, analog-to-digital conversion, frequency shifting, and EO instrumentation throughout the wider spectral range at which integrated lasers and photodetectors are now available in various photonics platforms \cite{piels2016,fang2006electrically,duan2014hybrid,Luo2015,shin2018band,li2019band,colucci2022unique,wen2022waveguide}.  

\textcolor{black}{Hybrid TFLN modulators can help integration and scalability of Pockels-effect modulators with silicon photonics. Our devices are based on low-temperature direct bonding of unetched TFLN to patterned and planarized Si waveguides, with the main fabrication steps described in Ref. \cite{mere2022modular}. For a specific MZ-EOM device design with short transition and phase-shift sections, we have shown that 110 GHz 3-dB EO bandwidth can be achieved while handling 110 mW of optical power  in the C-band \cite{valdez2022110}. Here we study a wider range of O-band and C-band devices systematically with different optical and RF designs, and explore strategies to reduce the half-wave voltage while also achieving high bandwidth in the hybrid bonded Si/LN platform. We design O-band and C-band 100-GHz-class MZ EOM devices using the same layer thicknesses of the Si layer, the TFLN layer and the thin oxide layer between the Si and the TFLN layers. The precise optical-RF index matching that is necessary for very high EO bandwidths ($>$100 GHz) was achieved in each wavelength band, firstly, by precisely tailoring the silicon waveguide width, and secondly, by adjusting the RF traveling wave electrode based on measurements of the layer thicknesses that are specific to each reticle of the wafer. Our approach results in integrated EO modulators on a Si photonics wafer which achieve $>$100 GHz EO bandwidth in both the O and C bands, a voltage-times-length figure-of-merit $V\pi L$ $<$ 3 V.cm in the C-band and 2 V.cm in the O-band (with device lengths up to $L=1$ cm), and an on-chip optical insertion loss (including the 3-dB couplers and phase-shifter sections) of less than 2 dB. These performance parameters lie well beyond what is possible using carrier-depletion MZ modulators in a traditional Si photonics platform \cite{witzens2018high} while offering easy integration with other Si photonic components through the continuation of the Si feeder waveguides which continue and extend without interruptions outside the bonded region \cite{wang2022monolithic,wang2022integrated}.} 

In Section 2, we describe the layer and device structure, and compare the EOM device designs for O-band and C-band operation.  The measurements of the EO parameters are described in Section 3. Section 4 further discusses some aspects of these hybrid Si-TFLN devices, followed by the conclusion in Section 5.

\section{Hybrid Design for O and C Band Operation}
\begin{figure}[ht]
\includegraphics[width=13.25cm]{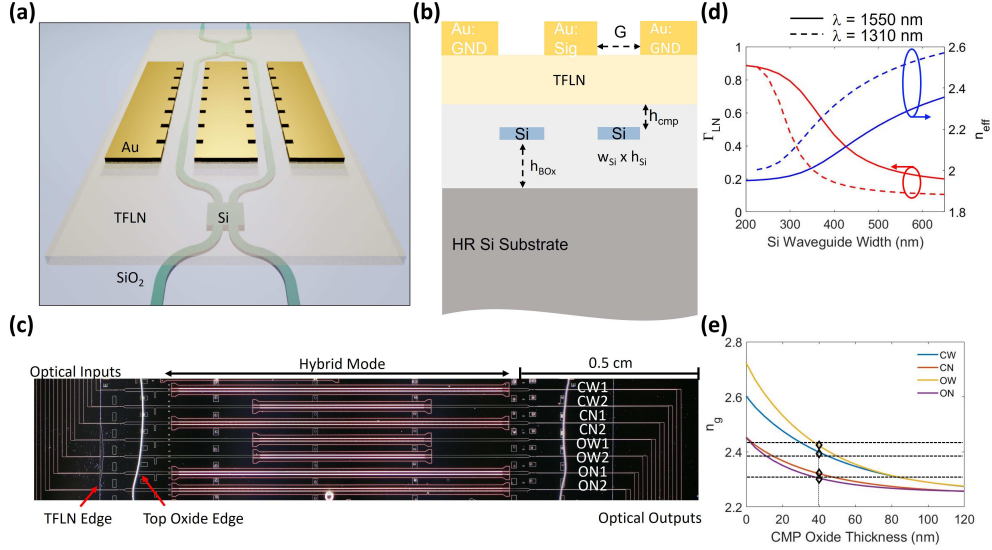}
\caption{(a) A perspective-view schematic (not to scale) of the hybrid bonded Si/LN MZM with slow-wave electrodes (SWEs). (b) A cross-section of the hybrid Si/LN MZM in the phase-shifter section, with a thin oxide layer (thickness $h_\mathrm{cmp}$) between the Si and LN layers. (c) An image of the fabricated chip with C-band (labeled Cxy) and O-band (labeled Oxy) MZM devices. (d) The simulated confinement factor of light in the LN film (left) and the effective index (right) of the fundamental $\mathrm{TE}_\mathrm{0}$ mode as a function of the Si waveguide width. (e) The simulated optical group index of the hybrid mode as a function of $h_\mathrm{cmp}$ for the four Si waveguide dimensions (Cx and Ox). The black diamonds mark the $h_\mathrm{cmp}$ of the fabricated devices. The black dashed lines correspond to the simulated RF index of the slow-wave electrodes.}
\label{fig1}
\end{figure}

\textcolor{black}{A perspective view of the MZM device is shown in Fig. \ref{fig1}(a). A simplified cross-section profile is shown schematically in Fig.\ref{fig1}(b), depicting only the layers which are necessary for modulation; additional optical waveguide layers, dopants, vias or metal layers may be included as part of the lower Si stack \cite{rusing2019toward,Boynton2020a}. Different Si waveguide widths are used in the input, transition, and phase shifter sections, and wider Si features can confine nearly all the optical power and allow low-loss optical transitions in the feeder waveguide across the bonded LN edges \cite{Weigel2016}. We could include both O-band and C-band designs in the same Si mask by thinning the Si layer from the standard 220 nm to 150 nm thickness. The thickness of this layer is nearly constant across the wafer. However, the thickness of the oxide layer, labeled “$h_\textrm{cmp}$” in Fig. \ref{fig1}(b), can vary across the wafer after chemical-mechanical polishing (CMP), as we have reported elsewhere \cite{Weigel2018e}. Here, $h_\textrm{cmp}$ is about 40 nm on average, but more accurate information is necessary to design the appropriate RF traveling-wave electrode structure on the layer of deposited Au on top of TFLN. Therefore, $h_\textrm{cmp}$ was measured in each reticle after CMP was completed, by performing ellipsometry on test features.}

Figure \ref{fig1}(c) shows a top-view darkfield optical microscope image of a fabricated hybrid bonded Si/LN chip. \textcolor{black}{The chip was fabricated using a hydrophilic bonding process with the process described in Ref.~\cite{mere2022modular}.} The edge couplers for fiber coupling are to the north-west and south-east edges (not shown in this reduced-area image). The transition from a mode that is highly-confined in Si (\textcolor{black}{650 nm wide Si waveguide}) to a mode in which light is mostly in TFLN (\textcolor{black}{$<$ 320 nm wide Si waveguide}) occurs under the bonded TFLN area. The TFLN layer thickness is about 580 nm, and it is transparent and cannot be easily seen in this image, but the perimeter of the bonded film is labeled “TFLN edge” in Fig. \ref{fig1}(c). An oxide cladding was deposited using a plasma-enhanced chemical vapor deposition (PECVD) process, and the oxide edge is shown in Fig. \ref{fig1}(c). Other relevant fabrication details are described in \cite{mere2022modular}. Multimode interference (MMI) waveguide couplers are used to split or combine the light into two equally weighted waveguides for the EO phase-shifter section. The MMI coupler width was 6 $\mu$m. The O- and C-band MZMs require different length couplers to maintain a high extinction ratio (ER). Based on simulations, the 3-dB 2x2 MMI lengths were designed to be 49 $\mu$m and 39 $\mu$m for the O- and C-band devices, respectively. The corners of the MMIs were angled to reduce optical reflections back to the source \cite{zhang2020broadband}. Note that in this case, as opposed to our previous work \cite{mere2022modular,wang2022monolithic}, the bonded LN film is over the MMI couplers \textcolor{black}{[see Figs. \ref{fig1}(a) and \ref{fig1}(c)]} and does not affect the coupling ratio due to the high optical confinement in the Si layer for such wide features. 

\textcolor{black}{After the MMI coupler, the Si waveguide width is adiabatically tapered down from the feeder section (650 nm width) to a smaller value (between 225 nm and 300 nm, depending on the design) for the phase-shifter section. This results in increasing the fraction of optical power that resides in the LN film. This effect is shown by the fraction ($\Gamma_\text{LN}$) of the integral over the cross-sectional profile of the Poynting power which resides in the LN region, as shown in Fig. \ref{fig1}(d). The Si waveguide width also affects the optical group index, $n_\textrm{g}$, of the hybrid mode, where the wider waveguide has a larger $n_\textrm{g}$ and this will matched by the RF electrode design. To compare their performance, we selected two Si waveguide widths for each wavelength band: The C-band modulators have a hybrid Si waveguide width of either 300 nm or 275 nm (labelled as devices \textcolor{black}{CW: C-band, wide; and CN: C-band, narrow, respectively}), while the O-band modulators have a width of 250 nm or 225 nm (labelled as devices \textcolor{black}{OW: O-band, wide; ON: O-band, respectively}).} 

As part of the wafer development processes, the wafers undergo CMP to achieve a planar surface for bonding. The CMP oxide thickness was measured across the wafer using ellipsometry and varies less than 2 nm in the bonded regions which are about 2 cm x 1 cm in size. In the bonded stack, the CMP oxide layer is between the Si waveguide and the bonded LN film and impacts $n_\textrm{g}$ as shown in Fig. \ref{fig1}(e). Thinner CMP oxide results in a tighter confinement of the optical mode to the Si waveguide, and increases $n_\textrm{g}$ and decreases the effective modal area. For the measured chip, the CMP oxide thickness is about 40 nm, resulting in a simulated $n_\textrm{g}$ of 2.38, 2.31, 2.43, and 2.30 for the CW, CN, OW, and ON devices respectively at 1550 nm and 1310 nm. 

High-speed EO modulation using traveling wave coplanar waveguides requires index matching between the RF and optical waves, ideally at all RF driving frequencies and optical wavelengths of interest.  \textcolor{black}{For both Si and LN, the optical refractive index increases as the wavelength of light decreases.} The optical group index of the hybrid mode was calculated as a function of chromatic (operating wavelength) and geometric dispersion (the cross-sectional parameters), namely the Si dimensions, TFLN thickness, and the amount of oxide between the Si strip and TFLN layer [see Fig. \ref{fig1}(e)]. \textcolor{black}{Two different widths of the Si waveguide in each wavelength band were chosen (OW, ON, CW, and CN), to study RF-optical index matching with either Si waveguide tapering or RF electrode design.} 

\begin{figure}[ht]
\includegraphics[width=4in,height=3.04in]{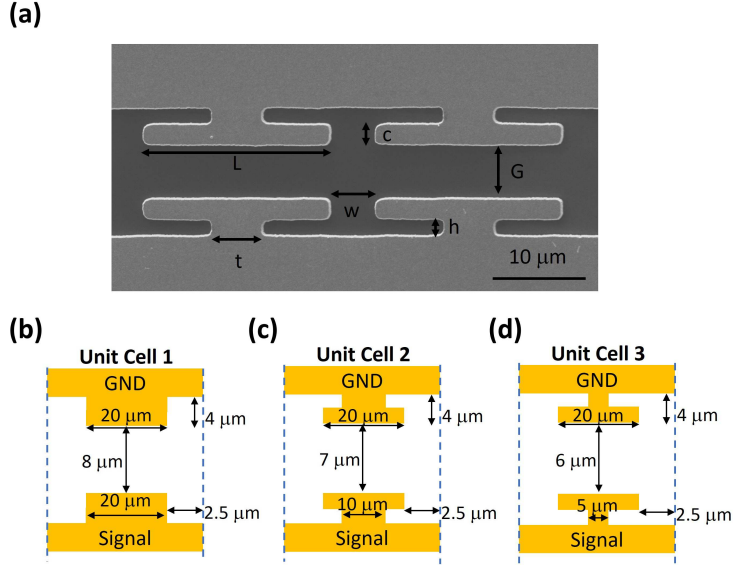}
\centering
\caption{(a) An SEM image of a T-rail SWE design used in this work with the SWE parameters labelled. (b) The top-view schematic of the periodic inductively loaded slot feature with slow-wave design parameters used for devices CN and ON. (c) The top-view schematic of the periodic capacitively loaded T-rail feature used for the CW devices. (d) The top-view schematic of the periodic capacitively loaded T-rail feature used for the OW devices. }

\label{fig2}
\end{figure}

Slow-wave electrodes (SWE) using inductively loaded or capacitively loaded elements were designed and fabricated to achieve RF-optical index matching between the O- and C-band hybrid optical modes and the RF coplanar waveguide (CPW) mode. SWE structures have been used to enable velocity matching of the electrodes to III-V semiconductor waveguides \cite{jaeger1992slow,Sakamoto1995}, and in LNOI modulators as well, particularly when using a low-RF-index substrate such as quartz \cite{kharel2021breaking,liu2021wideband} or for suspended/released modulators \cite{chen2022high}. SWEs are designed with either capacitively-loaded elements (T-rails), inductively loaded elements (slots), or a combination thereof to maintain velocity matching and impedance matching \cite{Sakamoto1995,spickermann1993millimetre,Rosa2018}. Figure \ref{fig2}(a) is an SEM image of one of the T-rail SWEs used in this work with the design parameters labelled. These additional parameters allow the RF dispersion (variation with frequency) to be adjusted. As the distance between the inner electrode edge and the inner T-rail edge increases [h in Fig \ref{fig2}(a)], the RF wave becomes slower which results in a larger RF effective index. Furthermore, the T-rail stem width [t, in Fig \ref{fig2}(a)] also affects the RF wave velocity, with narrower t corresponding to a slower RF field \cite{Sakamoto1995}. These three different SWE structures were designed to achieve index matching to the four different MZM designs as shown in Fig \ref{fig2}(b)-(d): a slot structure (L = t = 20 $\mu$m) for the CN and ON devices, a wider width T-rail for CW (L = 20 $\mu$m, t = 10 $\mu$m), and a narrow width T-rail for OW (L = 20 $\mu$m, t = 5 $\mu$m). The dashed lines in Fig \ref{fig1}(e) show the simulated RF indices, $n_\textrm{m}$, of the three SWEs at 110 GHz fabricated on this chip.

\textcolor{black}{The versatile electrode design offers a useful functionality in overcoming the effects of minor imperfections in the fabrication of the hybrid devices. The layer thicknesses were measured after planarization of the silicon and oxide layers was completed, and we observed some variations in layer thicknesses from die-to-die across the wafer. Before fabricating the electrodes, we also measured the optical transmission through an asymmetric Mach-Zehnder interferometer structure on the bonded chip, from which the optical refractive index of the target structure can be verified. As Figure \ref{fig1}(e) shows, the index matching condition can be achieved over a wide range of Si waveguide widths, CMP oxide thickness, or operational wavelengths by tuning the SWE parameters and tuning $n_\textrm{m}$ to match $n_\textrm{g}$. In this way, the electrode design was slightly altered for each pattern to trim devices and achieve the highest EO modulation bandwidths in each case.}

\section{Measurements}
\subsection{Electrical Characterization}
\begin{figure}[ht]
\includegraphics[width=\textwidth]{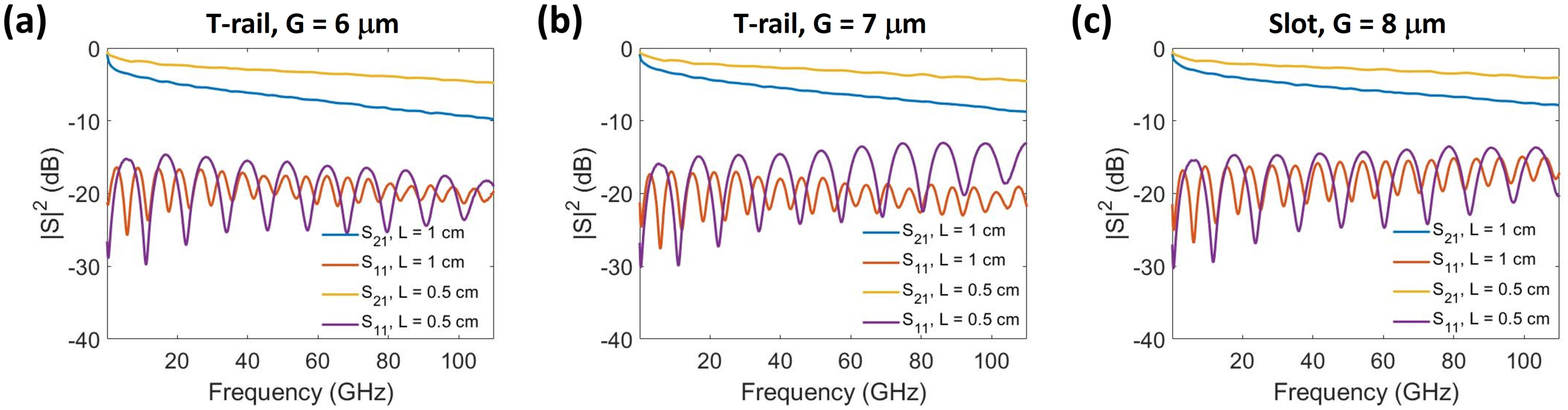}
\centering
\caption{Figure 3. The measured S-parameters of the 1.0 cm and 0.54 cm SWE with the following designs: (a) T-rail and G = 6 $\mu$m, (b) T-rail and G = 7 $\mu$m, (c) Slot and G = 8 $\mu$m. }
\label{fig3}
\end{figure}

The electrical response of the modulators was measured using a two-port Keysight PNA-X network analyzer with broadband frequency extenders, allowing for RF signals from 100 MHz up to 118 GHz to be applied to the SWEs. High speed ground-signal-ground (GSG) probes (FormFactor Infinity Probe) sourced and terminated the slow-wave transmission lines of the MZMs. Figures \ref{fig3}(a)-(c) show the measured electrical S-parameters of each of the three SWE designs (T-rail structure with G = 6 $\mu$m, T-rail structure with G = 7 $\mu$m, and slot structure with G = 8 $\mu$m) with both 1.0 cm and 0.54 cm length electrodes. The measured RF $|S_{21}|^2$ of the 1.0 cm long electrodes have a -6 dB drop at 75 GHz, while the 0.54 cm long electrodes have a -6 dB drop at greater than 118 GHz. 

\begin{figure}[ht]
\includegraphics[width=13.25cm]{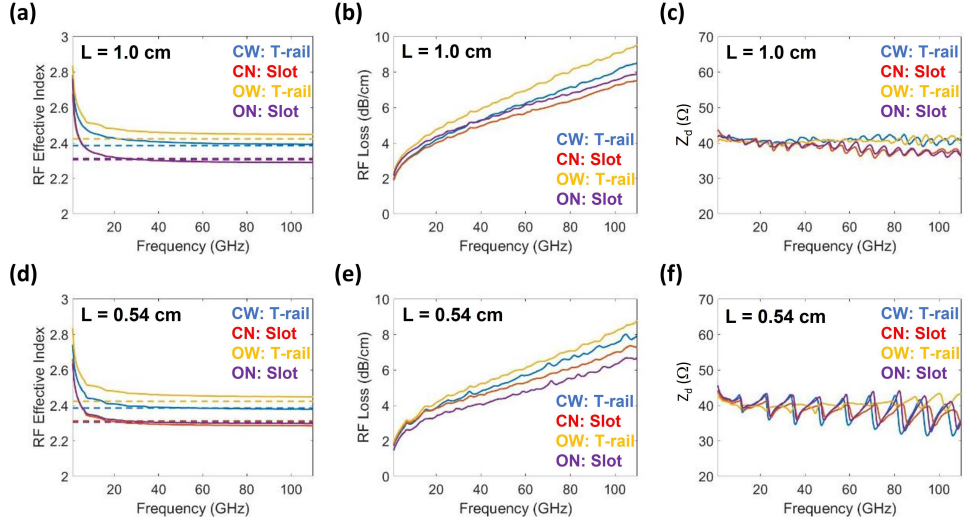}
\caption{The extracted RF characteristics of each of the SWE devices from the measured S-parameters. (a) RF effective index of the four 1.0 cm long SWEs. (b) RF propagation loss of each of the four 1.0 cm long SWEs. (c) RF driving impedance of each of the four 1.0 cm long SWEs. (d) RF effective index of the four 0.54 cm long SWE. (e) RF propagation loss of each of the four 0.54 cm long SWEs. (f) RF driving impedance of each of the four 0.54 cm long SWEs. The dashed lines in panels (a) and (d) correspond to the simulated $n_\textrm{g}$ for each of the C and O-band MZMs. }
\label{fig4}
\end{figure}

The RF index, loss, and impedance of the transmission lines extracted from the measured S-parameters are shown in Figs. \ref{fig4}(a) – (f) for each of the four electrode designs \cite{pozar2011microwave}. Figures \ref{fig4}(a) and \ref{fig4}(d) show that the SWE structures result in an RF-optical index mismatch of less than 1\% for both the C-band and O-band devices for both electrode lengths\textcolor{black}{, where the dashed lines correspond to the simulated $n_\textrm{g}$ of the four respective designs}. The RF propagation loss, $\alpha_\textrm{m}$, for each of the lines is less than 9 dB/cm at 110 GHz for all the structures [Fig. \ref{fig4}(b) and \ref{fig4}(e)], while the measured characteristic impedance, Zc, of the devices was about 40 $\Omega$ across the measured frequency range [Fig. \ref{fig4}(c) and \ref{fig4}(f)]. As discussed in section 4.2, this deviation from perfect 50 $\Omega$ impedance, which is characteristic of the source and detector, slightly lowers the measured EO 3-dB bandwidth. The RF back-reflection ($|S_{11}|^2$) is around -15 dB, and in some cases, around -20 dB as shown in Fig. \ref{fig3}. The skin-loss coefficient of the fabricated T-rail and Slot SWEs is 0.7-0.8  dB/cm-GHz$^{1/2}$, when the extracted RF loss is fit to the typical square-root frequency dependence equation. As there is less than 1\% index mismatch between the RF and optical traveling waves, the devices are RF loss limited. For these devices the gold electrodes are 0.75 $\mu$m thick. The RF losses can be reduced by increasing the electrode thickness via electroplating \cite{liu2022capacitively}. Alternatively, the Si substrate can be locally removed \cite{wang2022silicon}, or a lower loss material can be used as the substrate (such as quartz \cite{kharel2021breaking}); however, this will require further fabrication steps than needed here. 

\subsection{High-frequency EO response}
Light from separate O- and C-band instrument-grade lasers was edge coupled to the chip using a lensed fiber with a 4 $\mu$m nominal spot size. The laser wavelengths were set such that the asymmetric MZMs were biased at quadrature. The optical input power from the O- and C-band lasers was +12 dBm and +9 dBm, respectively. No optical amplifiers were used in the measurements. The propagation loss of the wide Si feeder sections and hybrid mode sections have been previously reported in the C-band as 0.8 dB/cm \cite{wang2022integrated} and 0.6 dB/cm \cite{Weigel2018e}, respectively. The wider feeder sections of the chip (only Si) that route from the input/output edge couplers to the hybrid phase-shifter section is 1.48 cm long; thus, 1.2 dB of the insertion loss is attributed to the feeder sections alone. The MMI couplers contribute to 0.2 dB of the total loss (per coupler), while there is about 0.1 dB of loss per LN edge. Inverse tapers with a resolution-limited tip width of 180 nm (tapering to 650 nm over 300 $\mu$m) were implemented at the input and output edges of the SOI chips to better match the lensed fiber mode to the waveguide modes in both wavelength bands. The optical loss attributed to the phase-shifter section of the device (hybrid mode along the electrode length) was estimated by comparing identical devices with different phase shifter lengths. For example, devices CN1 and CN2 are identical in optical design (Si waveguide, LN film) and RF design (SWE parameters), but the SWE length is 0.46 cm longer for CN1. Assuming all other losses are common, the loss from the phase-shifter would then be 1.5 dB/cm. The edge coupling losses are then estimated to be 4.2 dB per edge and 4.8 dB per edge for the C- and O-band, respectively. Although the waveguide taper tip width of 180 nm was limited by the resolution requirements of the foundry process, lower edge coupling losses can be achieved with smaller resolution or by using different types of edge couplers \cite{wang2016low,jia2018efficient,wang2019silicon,mu2020edge}. The insertion loss of the phase-shifter section, 3-dB MMI couplers, and LN transitions are thus calculated to be 1.6 dB and 2.1 dB for the 0.54 cm and 1.0 cm long phase-shifters, respectively. 

\begin{figure}[ht]
\includegraphics[width=13.25cm]{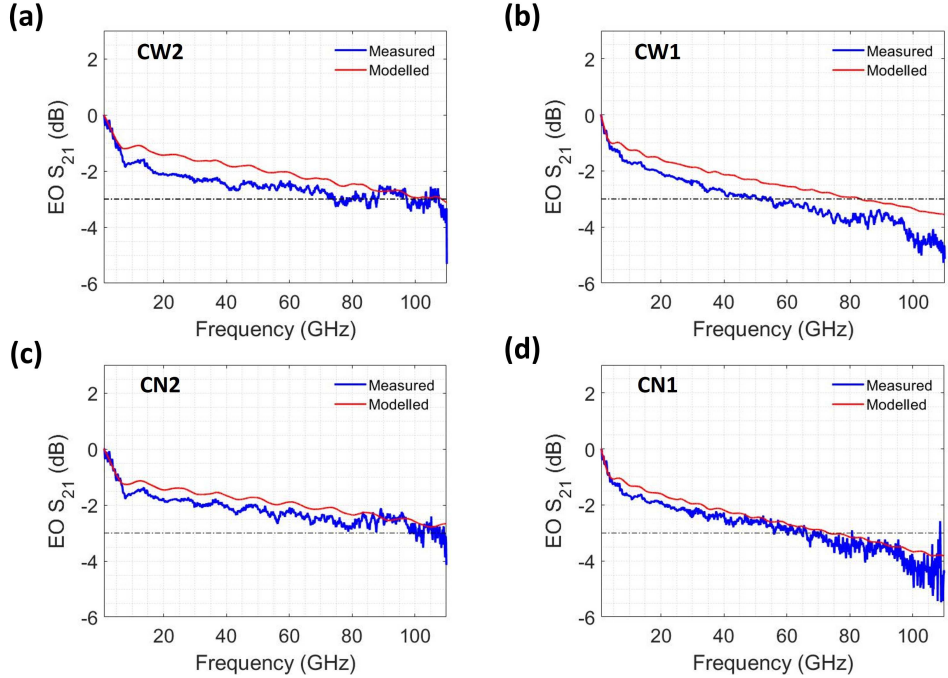}
\caption{The measured (blue) and modelled (red) EO responses of the hybrid bonded Si/LN MZMs designed for C-band. (a) T-rail SWE with G = 7 $\mu$m and L = 0.54 cm. (b) T-rail SWE with G = 7 $\mu$m and L = 1.0 cm. (c) Slot SWE with G = 8 $\mu$m and L = 0.54 cm.  (d) Slot SWE with G = 8 $\mu$m and L = 1.0 cm. }
\label{fig5}
\end{figure}

\begin{figure}[ht]
\includegraphics[width=13.25cm]{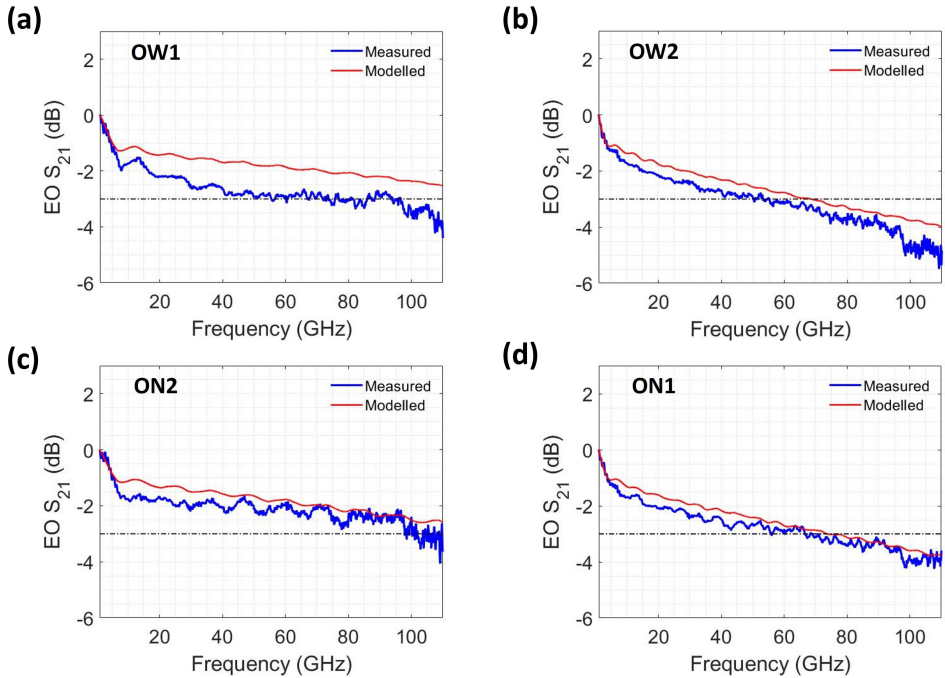}
\caption{The measured (blue) and modelled (red) EO responses of the hybrid bonded Si/LN MZMs designed for O-band. (a) T-rail SWE with G = 6 $\mu$m and L = 0.54 cm. (b) T-rail SWE with G = 6 $\mu$m and L = 1.0 cm.  (c) Slot SWE with G = 8 $\mu$m and L = 0.54 cm.  (d) Slot SWE with G = 8 $\mu$m and L = 1.0 cm.}
\label{fig6}
\end{figure}

To measure the EO performance, a 50 $\Omega$ load resistor was used to terminate the lines and a 110 GHz Keysight lightwave component analyzer (LCA) was used after full calibration of the probing setup. Figures \ref{fig5}(a)-(d) and \ref{fig6}(a)-(d) show the measured EO $S_{21}$ normalized to 1 GHz for the four C-band and four O-band MZMs, respectively. The red curves in Fig. \ref{fig5} and Fig. \ref{fig6} are the modelled EO responses for each device that was calculated using the simulated $n_\textrm{g}$ [dashed lines in Fig \ref{fig4}(a) and \ref{fig4}(d)], with the measured electrical RF characteristics ($n_\textrm{m}$, $\alpha_\textrm{m}$, and $Z_\textrm{c}$) from Figs. \ref{fig4}(a)-(f) in the following equations which are based on a traveling-wave model of the EO interaction \cite{ghione2009semiconductor}: 

\begin{subequations}
\begin{equation} \label{eq_EOR}
m(\omega)=\frac{R_{L}+R_{G}}{R_{L}}\left|\frac{Z_\mathrm{in}}{Z_\mathrm{in}+Z_\mathrm{G}}\right|\left|\frac{(Z_\mathrm{L}+Z_\mathrm{c})F(u_{+})+(Z_\mathrm{L}-Z_\mathrm{c})F(u_{-})}{(Z_\mathrm{L}+Z_\mathrm{c})\exp[\gamma_\mathrm{m}L]+(Z_\mathrm{L}-Z_\mathrm{c})\exp[-\gamma_{m}L]}\right|,
\end{equation}
\begin{equation}
Z_\mathrm{in}=Z_\mathrm{c}\frac{Z_\mathrm{L}+Z_\mathrm{c}\mathrm{tanh}(\gamma_\mathrm{m}L)}{Z_\mathrm{c}+Z_\mathrm{L}\mathrm{tanh}(\gamma_\mathrm{m}L)},
\end{equation}
\begin{equation} \label{eq_gm}
\gamma_\mathrm{m}=\alpha_\mathrm{m}+\frac{j\omega}{c}n_\mathrm{m}.
\end{equation}
\begin{equation} \label{eq_Fu}
F(u_{\pm}(P))=\frac{1-\exp[u_{\pm}]}{u_{\pm}}
\end{equation}
\begin{equation} \label{eq_upm}
u_{\pm}(P)=\pm\alpha_\mathrm{m}L+\frac{j\omega}{c}(\pm n_\mathrm{m}-n_\mathrm{g})L.
\end{equation}
\end{subequations}

where $R_{L,G}$ are the load and generator resistances, $Z_{\textrm{in},L,G}$ are the input, load, and generator impedances, and $\gamma_\textrm{m}$ is the complex propagation constant of the RF wave along the transmission line. 

The measured RF S-parameters from an electrical only measurement (Fig. \ref{fig3}) were used to model the EO response using Eq. (\ref{eq_EOR}). However, because the impedance of our SWE was not 50 $\Omega$, the electrical measurements show oscillations versus RF frequency [Figs. \ref{fig4}(c) and \ref{fig4}(f)] that result from the impedance mismatch from the 50-40-50 $\Omega$ system (Source-Transmission line-Termination) \cite{gopalakrishnan1994performance}. To approximately infer the “true” EO response, because the magnitude of the ripples is relatively small (much less than 0.5 dB), the frequency-dependent $Zc(f)$ data obtained from the PNA-X measurements was fit to a linear curve from 5 GHz to 110 GHz. Based on the agreement of the data and the simulated performance to within a fraction of a dB over a wide range of RF frequencies, we conclude that the EO model described by Eq. (\ref{eq_EOR}) together with this linear approximation captures the EO behavior adequately well. The initial drop-off at sub-10 GHz frequencies is due to a combination of skin-loss and the impedance mismatch to the 50 $\Omega$ source and load. The decaying oscillations that can be seen in Figs. \ref{fig5} and \ref{fig6} is also indicative of an impedance mismatch between the SWEs and the source and termination loads. Furthermore, the measured EO $S_{21}$ traces in Figs. \ref{fig5} and \ref{fig6} have a gentle slope and do not reach the -6 dB point (where the $V_\pi$ of the device would increase by a factor of two), nor the frequency cutoff regime. See section 4.1 and 4.2 below for more details. 

\subsection{Low-frequency $V_\pi L$}
\begin{figure}[ht]
\includegraphics[width=13.25cm]{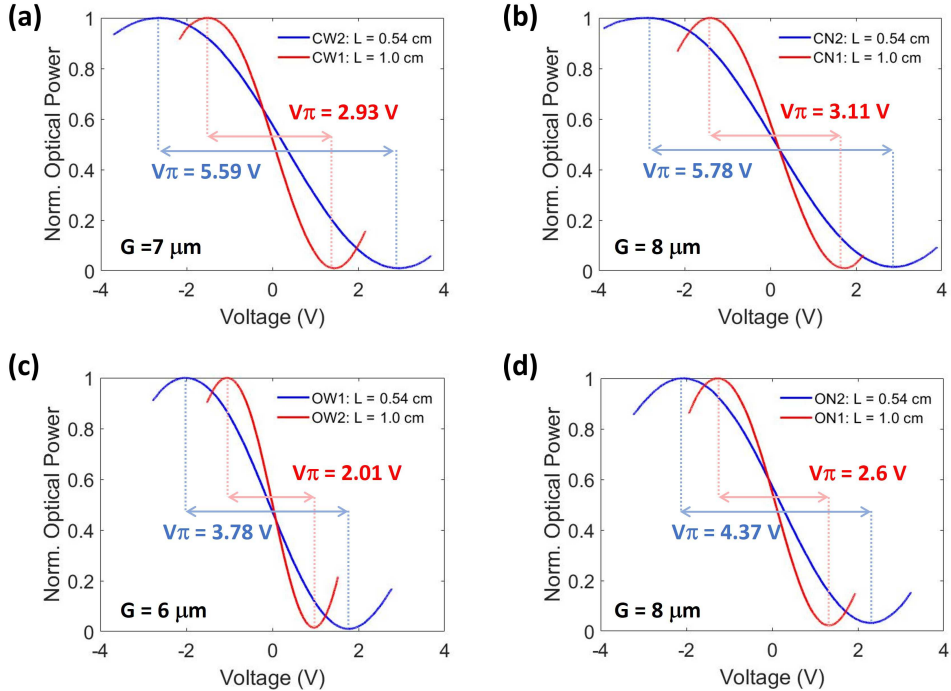}
\caption{The measured normalized optical power as a function of applied voltage to the hybrid bonded Si/LN MZMs for (a) CW MZMs with G = 7 $\mu$m.  (b) CN MZMs with G = 8 $\mu$m, (c) OW MZMs with G = 6 $\mu$m, (d) and ON MZMs with G = 8 $\mu$m. The blue traces correspond to the 0.54 cm long modulators and the red traces correspond to the 1.0 cm long modulators. }
\label{fig7}
\end{figure}

To measure the near-DC (1 kHz) half-wave voltage ($V_\pi$) of the O- and C-band modulators, these asymmetric MZMs were biased to quadrature by varying the laser wavelength. A waveform generator (Keysight 33600A) was used to apply 1 kHz sinusoidal signals to the devices, and the modulated optical signal was captured with a photodetector and an oscilloscope. The voltage levels applied to the modulators were set higher than the expected $V_\pi$ as to ensure that the full $\pi$ phase shift was induced. The waveform captured on the oscilloscope was post-processed using software to map the optical transmission to the applied voltage as shown in Fig. \ref{fig7}. From this relationship, a cosine-squared fit was used to evaluate $V_\pi$. The value of L is taken from the design: there are two phase-shifter interaction lengths of each of the four MZMs for L = 0.54 cm and 1.0 cm (the blue and red traces, respectively in Fig. \ref{fig7}). 

The half-wave voltage length product ($V_\pi L$) of an MZM using LN is given by the following equation
\begin{equation} \label{eqn_VpiL}
V_\pi L = \frac{n_\mathrm{eff}\lambda G}{2n_e^4 r_{33}\Gamma_\mathrm{mo}}
\end{equation}
where $n_\textrm{eff}$ is the effective refractive index of the optical mode, $\lambda$ is the operation wavelength (in vacuum), $G$ is the electrode gap distance between the ground and signal lines, $n_e$ is the extraordinary index of LN, $r_{33}$ is the linear Pockel’s coefficient in the crystal z-direction (30.8 pm/V), and $\Gamma_{mo}$ is the mode overlap integral between the traveling optical mode and RF mode. The factor of 2 in the denominator is included as the structure is driven in a push-pull configuration. Eq. (\ref{eqn_VpiL}) shows that the required driving voltage for a $\pi$ phase-shift reduces as the operation wavelength decreases. Furthermore, the effective area of the shorter wavelength optical modes is smaller, which implies that the electrode gap $G$ can be reduced without incurring high optical loss from optical absorption in the metal electrodes. Thus, the OW designs show the most efficient $V_\pi$ of the group of devices reported here, as shown in Fig. \ref{fig7}(c) with $V_\pi$ of 3.78 V and 2.01 V for the 0.54 cm and 1.0 cm long devices, respectively, which is a more efficient scaling of $V_\pi$ than would be predicted by the ratio of $\lambda$ at 1310 nm and 1550 nm.

When considering the overall system requirements to drive an EOM, it is important to know $V_\pi$ as a function of the modulation frequencies ($>$ 1 GHz), because the impact of RF loss, velocity mismatch, and impedance mismatch will also effect the needed voltage for a $\pi$ phase-shift \cite{howerton2002broadband}. The driving voltage as a function of modulation frequency is given by 
\begin{equation} \label{eqn_Vpif}
V_\pi(\omega) = V_\pi (DC)10^{-m(\omega)/20}
\end{equation}
where $m(\omega)$ is the measured normalized (for example, to 1 GHz) EO $S_{21}$. In this report, $V_\pi$(DC) is taken as the measured $V_\pi$ at 1 kHz as shown in Fig. \ref{fig6} for each MZM. \textcolor{black}{While a more accurate representation would be to normalize $m(\omega)$ to the same frequency as $V_\pi$(DC), this approximation is valid as the modulator responses are flat for frequencies less than 1 GHz (see Fig. \ref{fig11} below).} Figure \ref{fig8} shows the calculated RF $V_\pi$ which shows the effect of the SWE RF propagation loss, velocity, and impedance mismatch on the required driving voltage for the C and O-band MZMs. An increase of the $V_\pi$(DC) by a factor of $\sqrt{2}$ and $2$ correlate to the EO $S_{21}$ of the modulator decreasing by 3-dB and 6-dB, respectively. 

\begin{figure}[ht]
\includegraphics[width=13.25cm]{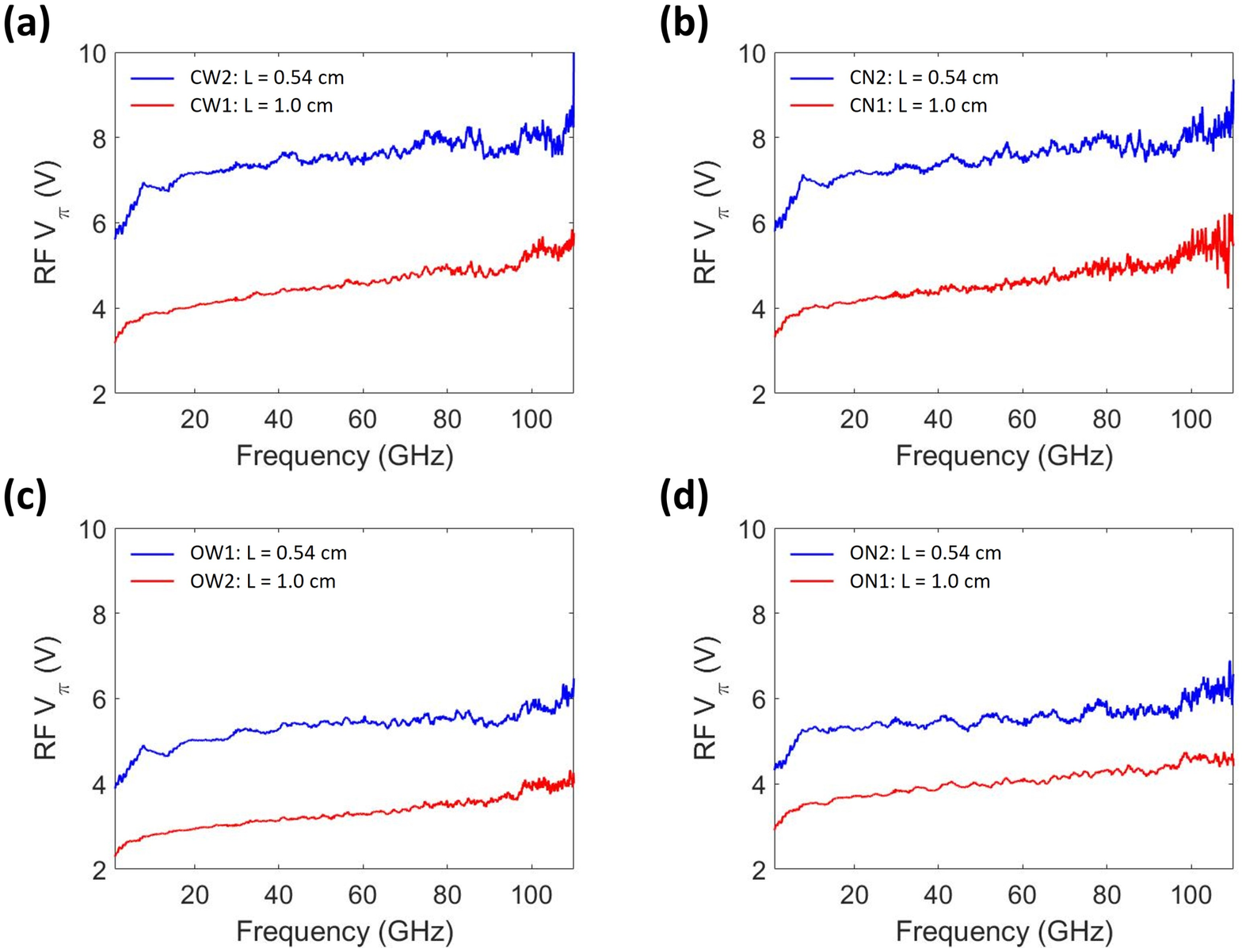}
\caption{The calculated RF $V_\pi$ as a function of driving frequency using Eq. (3) for the following Si/LN MZMs: (a) CW MZMs with G = 7 $\mu$m.  (b) CN MZMs with G = 8 $\mu$m, (c) OW MZMs with G = 6 $\mu$m, (d) and ON MZMs with G = 8 $\mu$m. The blue traces correspond to the 0.54 cm long modulators and the red traces correspond to the 1.0 cm long modulators. }
\label{fig8}
\end{figure}

\section{Discussion}
\textcolor{black}{Faced with the need to increase capacity, communications researchers are studying multiband optical networks which extend wavelength coverage beyond the C and L bands to other wavelength ranges such as the S, E and O bands \cite{9785841}. This requires new components, including switches and transceivers. Modulators in standard C-band transceivers show impaired performance when tested at other wavelengths. While mitigation using digital signal processing (DSP) compensation is being studied \cite{9723008,9591326}, this strategy adds to the cost and complexity. Complementary to software-based solutions are hardware (i.e., device) improvements, such as improving the modulator design. Lithium niobate modulators have been compared favorably to InP-based modulators for multiband operation, and TFLN modulators, in particular, have been identified as a potential future solution to the known limitations of current-generation LN modulators \cite{9723008,9591326}. However, these modulators are not yet mature, and many fundamental aspects of the new device platform are under study.} 

\textcolor{black}{Here, we showed that both O-band and C-band hybrid TFLN modulators can be made using the same TFLN layer thickness, the same Si layer thickness and with a single bonding step, with the only differences being in the silicon rib waveguide width, coupler design, and the customized electrodes, which are fabricated on top of the TFLN layer after bonding and handle removal. Within the design of each modulator, the MMI-based coupler is superior to directional-coupler designs for wider bandwidth, but customized designs were needed for the O-band and the C-band, and thus, the devices are not fully interchangeable, i.e., the C-band modulator is not ideal for O-band operation, or vice versa. Nevertheless, in our approach, the Si waveguide width is easily controlled, and the MMI couplers are designed in the Si layer, outside the bonded (hybrid) region, and they are not sensitive to the additional fabrication steps such as bonding and handle removal. When integrated with multiplexers, such hybrid modulators can easily cover wideband operation on the same chip, and deliver high-bandwidth and low-voltage performance in each band. A comparison between the different types of modulators studied here is presented in the following sections.} 

\subsection{O-band and C-band EO Device Comparison}
\begin{figure}[ht]
\includegraphics[width=\textwidth]{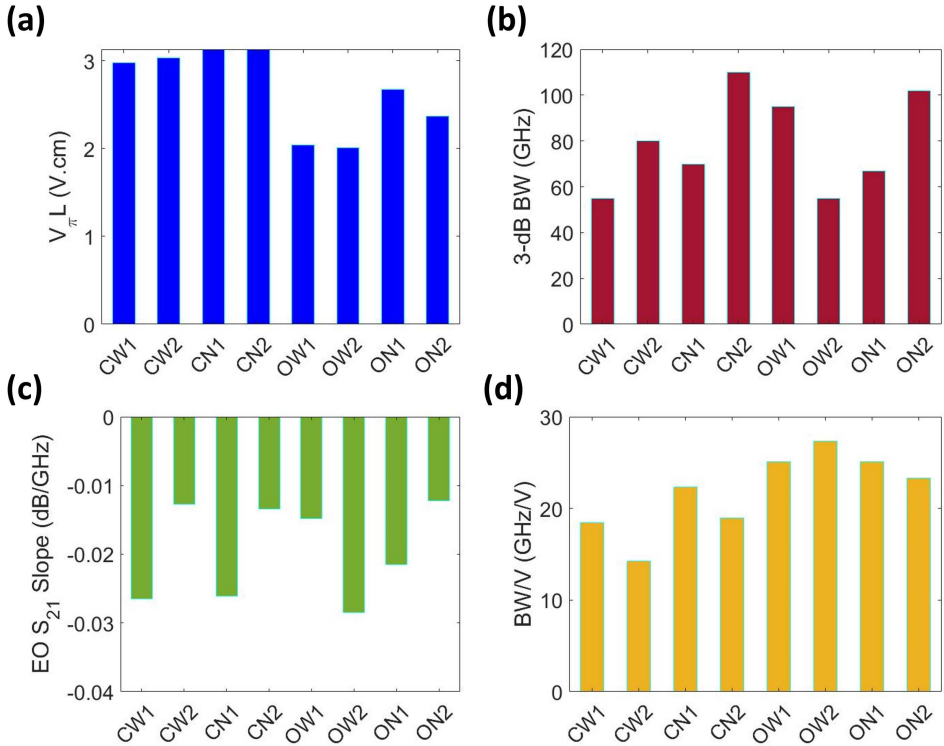}
\centering
\caption{\textcolor{black}{Summary bar graphs of the measured (a) $V_\pi L$, (b) 3-dB bandwidth (BW), (c) EO $S_{21}$ slope (from 10 GHz to 110 GHz), and (d) 3-dB BW-to-$V_\pi$ ratio for the hybrid bonded Si/LN C-band and O-band MZMs of the fabricated chip.} }
\label{fig9}
\end{figure}

Both the O-band and C-band devices show high bandwidth and low voltage. As shown in Fig. \ref{fig9}(a), the C-band modulators have a $V_\pi L$ of 2.9 V.cm to 3.1 V.cm, whereas the O-band modulators have a lower $V_\pi L$ of 2.0 V.cm to 2.3 V.cm. This is because the higher extraordinary index of refraction of LN, and the tighter optical mode confinement which allows for a reduced electrode gap at shorter wavelengths. The reduced gap has a more efficient EO effect, without incurring excess optical propagation loss from the proximity of the metal structures to the mode. These O-band devices can be further improved to have the Si waveguide width narrowed further and still maintain single mode operation and similar confinement factor ($\Gamma_\textrm{LN}$) as the C-band devices as shown in Fig. \ref{fig1}(e). 

The wide-Si modulators (CW and OW) have a hybrid optical mode which allows for an electrode gap spacing of G = 7 $\mu$m and 6 $\mu$m for CW and OW, respectively, instead of 8 $\mu$m for the narrow-Si modulators (CN and ON). The hybrid mode is more confined in the Si region than the LN film compared to the narrower designs (CN and ON), but the mode effective area is smaller. This tighter confinement allows for the electrode gap distance to be decreased, which also increases the modulation efficiency.

As seen in Fig.~\ref{fig9}(b), the 3-dB bandwidths of the 0.54~cm and 1.0~cm long MZMs are greater than 100 GHz and greater than 60 GHz, respectively for both O and C band designs. Note that although the initial 3-dB point of the wide Si waveguide 0.54 cm long MZMs (CW and OW in Fig. \ref{fig5}(a) and Fig. \ref{fig6}(a), respectively) occurs at 75 GHz, the response remains flat until 100 GHz with a slope of -0.013 dB/GHz (meaning a 1 dB drop over 100 GHz). An effective EOR slope from 10 GHz to 110 GHz is shown in Fig. \ref{fig9}(c) and ranges between -0.012 dB/GHz and -0.029 dB/GHz. This indicates that the modulators still have usable bandwidth up to, and probably higher than, 110 GHz. \textcolor{black}{Figure \ref{fig9}(d) summarizes the combined figure-of-merit, 3-dB bandwidth-to-$V_\pi$ ratio, of each modulator which shows the advantage of the O-band devices over the C-band counterparts. While both sets of modulators have been optimized for high-speed performance, the reduction of driving voltage for the O-band set allows for a higher $\mathrm{BW}/V_\pi$ which in turn reduces the power needed per high-speed modulation in a RF-photonic link. This may be helpful since communications in data centers and passive optical networks is energy-constrained.} 

\subsection{Discussion: RF Impedance}
\textcolor{black}{The drop in EO response by about 2 dB at low frequencies (around 1-10 GHz) that is seen in the EO measurements [see Fig. \ref{fig5} and \ref{fig6}] is attributed to the characteristic impedance of these traveling SWEs. Due to a design error, the impedance is around 40 $\Omega$ [see Fig. \ref{fig4}(d) and \ref{fig4}(f)], and is not matched to the source and load impedances, which are both 50 $\Omega$. The shorter devices (L = 0.54 cm) show this effect more clearly than the longer devices (L = 1.0 cm), where the additional RF loss reduces the back-reflection, and a similar effect has been seen in traditional LN modulators \cite{gopalakrishnan1994performance}. The impedance is sensitive to the electrode signal width to electrode gap ratio \cite{ghione2009semiconductor,honardoost2018high} and the mismatch in these devices can be corrected by redesigning the separation distance between the arms of the MZM. This would require a new round of lithography in the silicon layer followed by planarization, bonding, handle removal and electrode formation (i.e., a complete process flow). Alternatively, to avoid the cost of full refabrication, an on-chip termination can be fabricated on these devices to provide a customized matching condition to the SWE lines \cite{liu2022capacitively,huang2021high}. A smaller load impedance (for example, 30 $\Omega$) would result in peaking the response (when normalized to a certain frequency such as 1 GHz); however, this would be at the cost of causing larger back-reflections to the RF source, which is undesirable.}

\begin{figure}[ht]
\includegraphics[width=13.25cm]{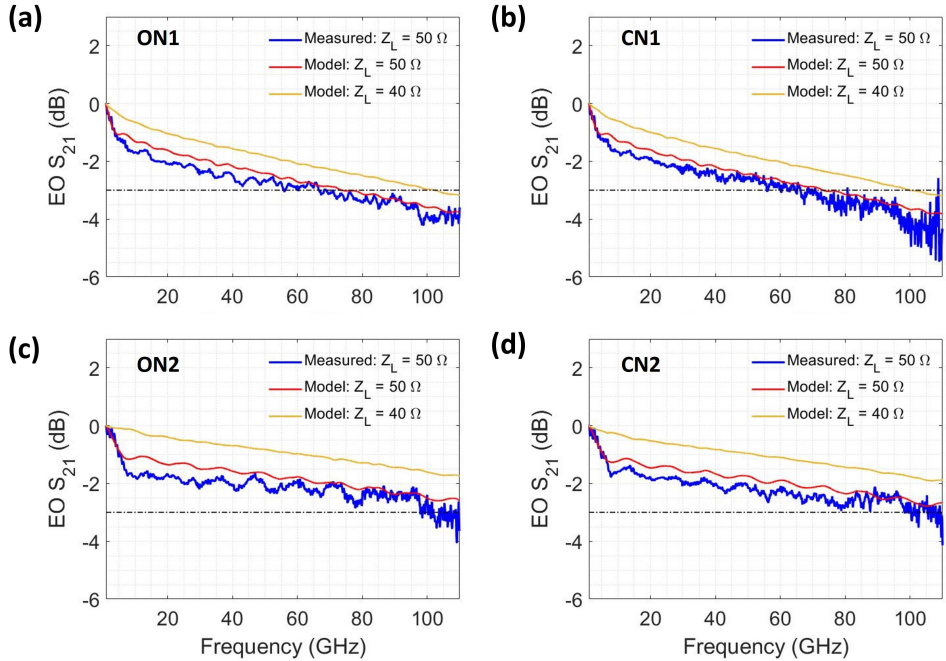}
\caption{The measured (blue) EO responses of the hybrid bonded Si/LN MZMs designed for: (a) O-band, Slot SWE with G = 8 $\mu$m and L = 1.0 cm. (b) C-band, Slot SWE with G = 8 $\mu$m and L = 1.0 cm.  (c) O-band, Slot SWE with G = 8 $\mu$m and L = 0.54 cm. (d) C-band, Slot SWE with G = 8 $\mu$m and L = 0.54 cm. The red and yellow curves correspond to the modelled EO response using Eq (1) and the measured RF characteristics in Fig \ref{fig3}(d)-(f), assuming $Z_\textrm{L}$ of 50 $\Omega$ and 40 $\Omega$, respectively.}
\label{fig10}
\end{figure}

\textcolor{black}{In the future, we believe that achieving a better match to 50 $\Omega$ source and termination loads, while maintaining the same levels of index matching and RF propagation loss that have already been achieved, should substantially improve the EO bandwidth. Figure \ref{fig10} is similar to the traces shown in Figs. \ref{fig5} and \ref{fig6}, with the addition of a yellow line which corresponds to the modelled EO responses with $Z_\textrm{L}$ = 40 $\Omega$ to match the characteristic impedance of the fabricated SWEs. In this case, the 3-dB EO bandwidths of the O- and C- band MZMs would be around 110 GHz for the 1.0 cm long modulators, and greater than 110 GHz for the 0.54 cm long modulators. This, in turn, will also decrease the RF $V_\pi$ (Fig. 8).} 

\textcolor{black}{Low frequency bias drift can affect x-cut LN modulators when there is a buffer oxide layer between the driving electrodes and LN surface~\cite{yamada1981dc,gee1985minimizing,nagata1993temperature,salvestrini2011analysis}. There is no such oxide layer in our devices since the gold (Au) electrodes are patterned directly on the LN surface. While we do not know if surface charge accumulation plays a significant role, we think it is unlikely as there are large Au ground planes in contact with the surface. In our measurements, we did not use a DC bias but instead, tuned the wavelength with a tunable laser, as each of the modulators has an asymmetric path-length difference. Long-term drift measurements and compensation using a bias controller will be studied in the future. We have observed that the measured value of $V_\pi$ is constant in the range of 0.1 to 10 MHz \cite{valdez2022110}. The simulation model fits the measurements well, when assuming that the transmission line impedance is 40 $\Omega$ system, as shown in Fig. \ref{fig11}, and the measured and modelled responses are flat from 0.1 to 1 GHz.}

\subsection{Discussion: Normalization of EO Response}
\begin{figure}[ht]
\includegraphics[width=\textwidth]{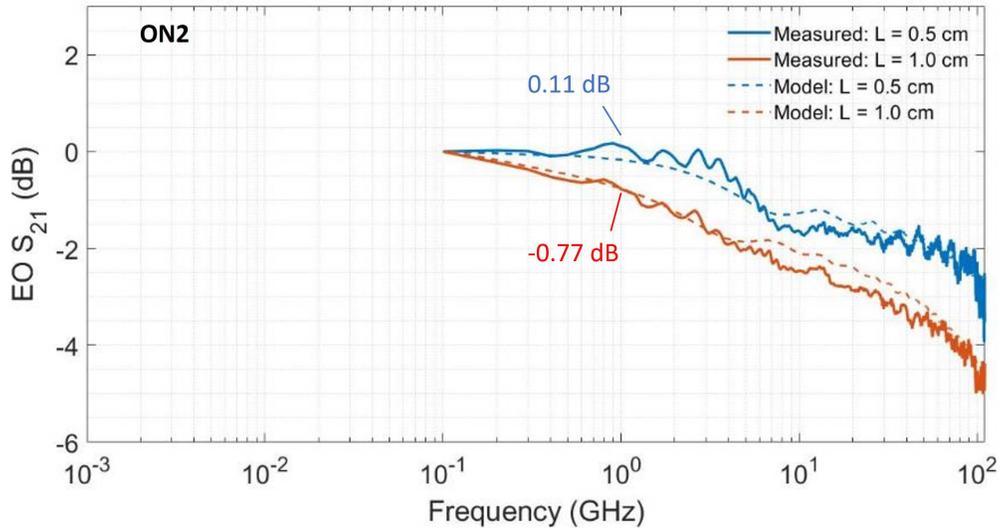}
\caption{The measured (solid) and modelled (dashed) EO response normalized to 0.1 GHz for the ON devices with 8 $\mu$m gap slot SWEs. }
\label{fig11}
\end{figure}

\textcolor{black}{The EO response was measured over 100 MHz to 110 GHz using the available range of the LCA instrument. The actual high frequency behavior of $|m(\omega)|$ is not affected by what value of $\omega$ is chosen for the denominator in Eq. (\ref{eq_EOR}) Since the actual value of $|m(\omega=0)|$ in Eq. (\ref{eq_EOR}) is not known, it is usually taken as the value at 1 GHz \cite{mere2022modular,wang2022monolithic,kharel2021breaking,nelan2022integrated} though values as high as 5 GHz have also been used \cite{alam2022net}.  For the devices (ON1 and ON2) shown in Fig. \ref{fig11}, the EO response changes by +0.11 dB from 100 MHz to 1 GHz for an L = 0.54 cm device, and by -0.77 dB for an L = 1.0 cm device. Often-quoted parameters such as the 3-dB bandwidth do depend on the reference point, which should therefore be clearly stated. Since the half-wave voltage $V_\pi$ can be measured down to very low frequencies, it is convenient to plot $V_\pi(\omega)$ as shown in Fig. \ref{fig8}, which shows a direct relationship to both $m(\omega)$ through Eq. (\ref{eqn_Vpif}), and graph the trend all the way from near DC to the highest modulation frequencies measured by the LCA. }

\section{Conclusion}
In conclusion, we have demonstrated that a hybrid bonded Si/TFLN platform can be used for high-speed and low voltage EOMs in both the O and C wavelength bands using the same fabrication process, a common layer stack, and on the same chip. Standard Si manufacturing processes were used to define the optical routing, splitting, and tapering of the waveguides. The top oxide layer of the SOI wafer was chemical-mechanically polished to provide a bondable surface and no etching or patterning of the LN layer is required. We show that the velocity matching condition for high-speed modulation can be precisely achieved \textcolor{black}{(to less than 1\%)} by changing the Si waveguide width (tuning $n_\textrm{g}$) and the parameters of the travelling slow-wave electrode structures (tuning $n_\textrm{m}$) \textcolor{black}{across different wavelength bands to account for both chromatic dispersion and local variations in geometry}. Four designs were reported here, two in the O-band and two in the C-band, which differ only in the widths of the features in Si layer and in the traveling slow-wave electrode designs. The high-frequency EO response of the 0.54 cm and 1.0 cm long devices cross the 3 dB line (referenced to 1 GHz) at about 100 GHz and 60 GHz, respectively, for both O and C bands. However, a single number such as 3 dB bandwidth may not fully capture the device performance, since the gentle slopes of the modulator response, and the fact that the EO response has not reached the cut-off frequency regime indicates useful EO bandwidth beyond 110 GHz in both cases. \textcolor{black}{While all eight reported modulators are shown to have high electro-optic modulation bandwidth, the O-band devices prove to have a higher modulation efficiency, and the combination of high bandwidth and low voltage can greatly benefit short-range communications in data centers.}



\section*{Funding} 
NASA (80NSSC17K0166), ONR (N00014-21-1-2805), DOD (HR001120S0008), U.S. Government.
\section*{Acknowledgments}
The authors thank: M. R\"using, P. O. Weigel and J. Zhao (formerly of UC San Diego) for earlier contributions and discussions on this topic; A. Lentine, N. Boynton, T. A. Friedman, S. Arterburn, C. Dallo, A. T. Pomerene, A. L. Starbuck, and D. C. Trotter (Sandia National Laboratories) for discussions and fabrication assistance; C. Coleman, R. Scott, B. Szafraniec, G. Vanwiggeren, V. Moskalenko, K.K. Abdelsalam and G. Lee (Keysight Technologies) for discussions and measurement assistance. Part of this work was performed at the San Diego Nanotechnology Infrastructure (SDNI) of UCSD, a member of the National Nanotechnology Coordinated Infrastructure, which is supported by the National Science Foundation (Grant ECCS-2025752). This research was developed in part with funding from the Defense Advanced Research Projects Agency (DARPA) and the U.S. Government. This paper describes objective technical results and analysis. The views, opinions and/or findings expressed are those of the authors alone and should not be interpreted as representing the official views or policies of the Department of Defense or the U.S. Government.

\section*{Disclosures}
The authors declare no conflicts of interest.

\section*{Data availability.} Data underlying the results presented in this paper are not publicly available at this time but may be obtained from the authors upon reasonable request.


\bibliographystyle{unsrt} 
\bibliography{References_O-C_SiLN_MZM_Rev3_arxiv}

\begin{thebibliography}{10}

\bibitem{sohler2008integrated}
Wolfgang Sohler, Hui Hu, Raimund Ricken, Viktor Quiring, Christoph Vannahme,
  Harald Herrmann, Daniel B{\"u}chter, Selim Reza, Werner Grundk{\"o}tter,
  Sergey Orlov, Hubertus Suche, Rahman Nouroozi, and Yoohong Min.
\newblock Integrated optical devices in lithium niobate.
\newblock {\em Optics and Photonics News}, 19(1):24--31, 2008.

\bibitem{Boes2018}
Andreas Boes, Bill Corcoran, Lin Chang, John Bowers, and Arnan Mitchell.
\newblock {Status and Potential of Lithium Niobate on Insulator (LNOI) for
  Photonic Integrated Circuits}.
\newblock {\em Laser \& Photonics Reviews}, 12(4):1700256, feb 2018.

\bibitem{wang2018integrated}
Cheng Wang, Mian Zhang, Xi~Chen, Maxime Bertrand, Amirhassan Shams-Ansari,
  Sethumadhavan Chandrasekhar, Peter Winzer, and Marko Lon{\v{c}}ar.
\newblock {Integrated lithium niobate electro-optic modulators operating at
  CMOS-compatible voltages}.
\newblock {\em Nature}, 562(7725):101--104, 2018.

\bibitem{he2019high}
Mingbo He, Mengyue Xu, Yuxuan Ren, Jian Jian, Ziliang Ruan, Yongsheng Xu,
  Shengqian Gao, Shihao Sun, Xueqin Wen, Lidan Zhou, Lin Liu, Changjian Guo,
  Hui Chen, Siyuan Yu, Liu Liu, and Xinlun Cai.
\newblock {High-performance hybrid silicon and lithium niobate Mach-Zehnder
  modulators for 100 Gbit s$^{-1}$ and beyond}.
\newblock {\em Nature Photonics}, 13(5):359--364, 2019.

\bibitem{zhu2021integrated}
Di~Zhu, Linbo Shao, Mengjie Yu, Rebecca Cheng, Boris Desiatov, CJ~Xin, Yaowen
  Hu, Jeffrey Holzgrafe, Soumya Ghosh, Amirhassan Shams-Ansari, Eric Puma, Neil
  Sinclair, Christian Reimer, Mian Zhang, and Marko Loncar.
\newblock Integrated photonics on thin-film lithium niobate.
\newblock {\em Advances in Optics and Photonics}, 13(2):242--352, 2021.

\bibitem{IEEE2015}
Physical layer specifications and management parameters for 40 gb/s and 100
  gp/s operation over fiber optic cables.
\newblock {\em IEEE Standard for Ethernet}.

\bibitem{stenger2013engineered}
Vincent Stenger, James Toney, Andrea Pollick, James Busch, Jon Scholl, Peter
  Pontius, and Sri Sriram.
\newblock Engineered thin film lithium niobate substrate for high
  gain-bandwidth electro-optic modulators.
\newblock In {\em CLEO: Science and Innovations}, pages CW3O--3. Optica
  Publishing Group, 2013.

\bibitem{sun2020hybrid}
Shihao Sun, Mingbo He, Siyuan Yu, and Xinlun Cai.
\newblock Hybrid silicon and lithium niobate {Mach-Zehnder} modulators with
  high bandwidth operating at {C-band and O-band}.
\newblock In {\em CLEO: Science and Innovations}, pages STh1F--4. Optica
  Publishing Group, 2020.

\bibitem{piels2016}
Molly Piels and John~E. Bowers.
\newblock {\em {Photodetectors for silicon photonic integrated circuits}}.
\newblock Elsevier Ltd, 2016.

\bibitem{fang2006electrically}
Alexander~W Fang, Hyundai Park, Oded Cohen, Richard Jones, Mario~J Paniccia,
  and John~E Bowers.
\newblock Electrically pumped hybrid {AlGaInAs-silicon} evanescent laser.
\newblock {\em Optics Express}, 14(20):9203--9210, 2006.

\bibitem{duan2014hybrid}
Guang-Hua Duan, Christophe Jany, Alban Le~Liepvre, Alain Accard, Marco Lamponi,
  Dalila Make, Peter Kaspar, Guillaume Levaufre, Nils Girard, Fran{\c{c}}ois
  Lelarge, Jean-Marc Fedeli, Antoine Descos, Badhise~Ben Bakir, Sonia
  Messaoudene, Damien Bordel, Sylvie Menezo, Guilhem~de Valicourt, Shahram
  Keyvaninia, Gunther Roelkens, Dries~Van Thourhout, Davd~J. Thomson,
  Frederic~Y. Gardes, and Graham~T. Reed.
\newblock Hybrid {III--V} on {Silicon Lasers for Photonic Integrated Circuits
  on Silicon}.
\newblock {\em IEEE Journal of Selected Topics in Quantum Electronics},
  20(4):158--170, 2014.

\bibitem{Luo2015}
Xianshu Luo, Yulian Cao, Junfeng Song, Xiaonan Hu, Yuanbing Cheng, Chengming
  Li, Chongyang Liu, Tsung~Yang Liow, Mingbin Yu, Hong Wang, Qi~Jie Wang, and
  Patrick Guo~Qiang Lo.
\newblock {High-throughput multiple dies-to-wafer bonding technology and
  III/V-on-Si hybrid lasers for heterogeneous integration of optoelectronic
  integrated circuits}.
\newblock {\em Frontiers in Materials}, 2(April):1--21, 2015.

\bibitem{shin2018band}
Dongjae Shin, Jungho Cha, Sunggu Kim, Yongwhak Shin, Kwansik Cho, Kyoungho Ha,
  Gitae Jeong, Hyeongsun Hong, Kyupil Lee, and Ho-Kyu Kang.
\newblock O-band {DFB} laser heterogeneously integrated on a bulk-silicon
  platform.
\newblock {\em Optics Express}, 26(11):14768--14774, 2018.

\bibitem{li2019band}
Keshuang Li, Zizhuo Liu, Mingchu Tang, Mengya Liao, Dongyoung Kim, Huiwen Deng,
  Ana~M Sanchez, R~Beanland, Mickael Martin, Thierry Baron, Siming Chen, Jiang
  Wu, Alwyn Seeds, and Huiyan Liu.
\newblock O-band {InAs/GaAs} quantum dot laser monolithically integrated on
  exact (0 0 1) {Si} substrate.
\newblock {\em Journal of Crystal Growth}, 511:56--60, 2019.

\bibitem{colucci2022unique}
Davide Colucci, Marina Baryshnikova, Yuting Shi, Yves Mols, Muhammad Muneeb,
  Yannick De~Koninck, Didit Yudistira, Marianna Pantouvaki, Joris
  Van~Campenhout, Robert Langer, Dries~Van Thourhout, and Bernardette Kunert.
\newblock Unique design approach to realize an {O-band} laser monolithically
  integrated on 300 mm {Si} substrate by nano-ridge engineering.
\newblock {\em Optics Express}, 30(8):13510--13521, 2022.

\bibitem{wen2022waveguide}
Pengyan Wen, Preksha Tiwari, Svenja Mauthe, Heinz Schmid, Marilyne Sousa,
  Markus Scherrer, Michael Baumann, Bertold~Ian Bitachon, Juerg Leuthold, Bernd
  Gotsmann, and Kirsten~E. Moselund.
\newblock Waveguide coupled {III-V} photodiodes monolithically integrated on
  {Si}.
\newblock {\em Nature Communications}, 13(1):1--11, 2022.

\bibitem{mere2022modular}
Viphretuo Mere, Forrest Valdez, Xiaoxi Wang, and Shayan Mookherjea.
\newblock A modular fabrication process for thin-film lithium niobate
  modulators with silicon photonics.
\newblock {\em J.Phys. Photonics}, 4(2):024001, 2022.

\bibitem{valdez2022110}
Forrest Valdez, Viphretuo Mere, Xiaoxi Wang, Nicholas Boynton, Thomas~A
  Friedmann, Shawn Arterburn, Christina Dallo, Andrew~T Pomerene, Andrew~L
  Starbuck, Douglas~C Trotter, Anthony~L Lentine, and Shayan Mookherjea.
\newblock 110 {GHz}, 110 {mW} hybrid silicon-lithium niobate {Mach-Zehnder}
  modulator.
\newblock {\em Scientific Reports}, 12(1):1--11, 2022.

\bibitem{witzens2018high}
Jeremy Witzens.
\newblock High-speed silicon photonics modulators.
\newblock {\em Proceedings of the IEEE}, 106(12):2158--2182, 2018.

\bibitem{wang2022monolithic}
Xiaoxi Wang, Forrest Valdez, Viphretuo Mere, and Shayan Mookherjea.
\newblock {Monolithic Integration of 110 GHz Thin-film Lithium Niobate
  Modulator and High-Q Silicon Microring Resonator for Photon-Pair Generation}.
\newblock In {\em 2022 Conference on Lasers and Electro-Optics (CLEO)}, pages
  1--2. IEEE, 2022.

\bibitem{wang2022integrated}
Xiaoxi Wang, Forrest Valdez, Viphretuo Mere, and Shayan Mookherjea.
\newblock Integrated thin-silicon passive components for hybrid silicon-lithium
  niobate photonics.
\newblock {\em Optics Continuum}, 1(10):2233--2244, 2022.

\bibitem{rusing2019toward}
Michael Rusing, Peter~O Weigel, Jie Zhao, and Shayan Mookherjea.
\newblock Toward 3d integrated photonics including lithium niobate thin films:
  a bridge between electronics, radio frequency, and optical technology.
\newblock {\em IEEE Nanotechnology Magazine}, 13(4):18--33, 2019.

\bibitem{Boynton2020a}
Nicholas Boynton, Hong Cai, Michael Gehl, Shawn Arterburn, Christina Dallo,
  Andrew Pomerene, Andrew Starbuck, Dana Hood, Douglas~C. Trotter, Thomas
  Friedmann, Christopher~T. DeRose, and Anthony Lentine.
\newblock {A heterogeneously integrated silicon photonic/lithium niobate
  travelling wave electro-optic modulator}.
\newblock {\em Optics Express}, 28(2):1868--1884, 2020.

\bibitem{Weigel2016}
Peter~O. Weigel, Marc Savanier, Christopher~T. Derose, Andrew~T. Pomerene,
  Andrew~L. Starbuck, Anthony~L. Lentine, Vincent Stenger, and Shayan
  Mookherjea.
\newblock {Lightwave Circuits in Lithium Niobate through Hybrid Waveguides with
  Silicon Photonics}.
\newblock {\em Scientific Reports}, 6(November 2015):1--9, 2016.

\bibitem{Weigel2018e}
Peter~O. Weigel, Jie Zhao, Kelvin Fang, Hasan Al-Rubaye, Douglas Trotter, Dana
  Hood, John Mudrick, Christina Dallo, Andrew~T. Pomerene, Andrew~L. Starbuck,
  Christopher~T. DeRose, Anthony~L. Lentine, Gabriel Rebeiz, and Shayan
  Mookherjea.
\newblock {Bonded thin film lithium niobate modulator on a silicon photonics
  platform exceeding 100 GHz 3-dB electrical modulation bandwidth}.
\newblock {\em Optics Express}, 26(18):23728, 2018.

\bibitem{zhang2020broadband}
Jin Zhang, Liangshun Han, Bill Ping-Piu Kuo, and Stojan Radic.
\newblock Broadband angled arbitrary ratio {SOI} {MMI} couplers with enhanced
  fabrication tolerance.
\newblock {\em Journal of Lightwave Technology}, 38(20):5748--5755, 2020.

\bibitem{jaeger1992slow}
NAF Jaeger and Zachary~KF Lee.
\newblock Slow-wave electrode for use in compound semiconductor electrooptic
  modulators.
\newblock {\em IEEE Journal of Quantum Electronics}, 28(8):1778--1784, 1992.

\bibitem{Sakamoto1995}
S.~R. Sakamoto, R.~Spickermann, and N.~Dagli.
\newblock {Narrow gap coplanar slow wave electrode for travelling wave
  electro-optic modulators}.
\newblock {\em Electronics Letters}, 31(14):1183--1185, 1995.

\bibitem{kharel2021breaking}
Prashanta Kharel, Christian Reimer, Kevin Luke, Lingyan He, and Mian Zhang.
\newblock Breaking voltage--bandwidth limits in integrated lithium niobate
  modulators using micro-structured electrodes.
\newblock {\em Optica}, 8(3):357--363, 2021.

\bibitem{liu2021wideband}
Xuecheng Liu, Bing Xiong, Changzheng Sun, Jian Wang, Zhibiao Hao, Lai Wang,
  Yanjun Han, Hongtao Li, Jiadong Yu, and Yi~Luo.
\newblock Wideband thin-film lithium niobate modulator with low
  half-wave-voltage length product.
\newblock {\em Chinese Optics Letters}, 19(6):060016, 2021.

\bibitem{chen2022high}
Gengxin Chen, Kaixuan Chen, Ranfeng Gan, Ziliang Ruan, Zong Wang, Pucheng
  Huang, Chao Lu, Alan Pak~Tao Lau, Daoxin Dai, Changjian Guo, and Liu Liu.
\newblock High performance thin-film lithium niobate modulator on a silicon
  substrate using periodic capacitively loaded traveling-wave electrode.
\newblock {\em APL Photonics}, 7(2):026103, 2022.

\bibitem{spickermann1993millimetre}
R~Spickermann and N~Dagli.
\newblock {Millimetre wave coplanar slow wave structure on GaAs suitable for
  use in electro-optic modulators}.
\newblock {\em Electronics Letters}, 29(9):774--775, 1993.

\bibitem{Rosa2018}
{\'{A}}lvaro Rosa, Steven Verstuyft, Antoine Brimont, Dries~Van Thourhout, and
  Pablo Sanchis.
\newblock {Microwave index engineering for slow-wave coplanar waveguides}.
\newblock {\em Scientific Reports}, 8(1):1--8, 2018.

\bibitem{pozar2011microwave}
David~M Pozar.
\newblock {\em Microwave engineering}.
\newblock John Wiley \& Sons, 2011.

\bibitem{liu2022capacitively}
Xuecheng Liu, Bing Xiong, Changzheng Sun, Zhibiao Hao, Lai Wang, Jian Wang,
  Yanjun Han, Hongtao Li, and Yi~Luo.
\newblock Capacitively-loaded thin-film lithium niobate modulator with
  ultra-flat frequency response.
\newblock {\em IEEE Photonics Technology Letters}, 34(16):854--857, 2022.

\bibitem{wang2022silicon}
Zong Wang, Gengxin Chen, Ziliang Ruan, Ranfeng Gan, Pucheng Huang, Zhiwen
  Zheng, Liwang Lu, Jun Li, Changjian Guo, Kaixuan Chen, and Liu Liu.
\newblock {Silicon--Lithium Niobate Hybrid Intensity and Coherent Modulators
  Using a Periodic Capacitively Loaded Traveling-Wave Electrode}.
\newblock {\em ACS Photonics}, 9(8):2668--2675, 2022.

\bibitem{wang2016low}
Jing Wang, Yi~Xuan, Chunghun Lee, Ben Niu, Lei Liu, Gordon~Ning Liu, and
  Minghao Qi.
\newblock Low-loss and misalignment-tolerant fiber-to-chip edge coupler based
  on double-tip inverse tapers.
\newblock In {\em Optical Fiber Communication Conference}, pages M2I--6. Optica
  Publishing Group, 2016.

\bibitem{jia2018efficient}
Lianxi Jia, Chao Li, Tsung-Yang Liow, and Guo-Qiang Lo.
\newblock Efficient suspended coupler with loss less than- 1.4 {dB} between
  {Si-photonic} waveguide and cleaved single mode fiber.
\newblock {\em Journal of Lightwave Technology}, 36(2):239--244, 2018.

\bibitem{wang2019silicon}
Xiaodong Wang, Xueling Quan, Min Liu, and Xiulan Cheng.
\newblock Silicon-nitride-assisted edge coupler interfacing with high numerical
  aperture fiber.
\newblock {\em IEEE Photonics Technology Letters}, 31(5):349--352, 2019.

\bibitem{mu2020edge}
Xin Mu, Sailong Wu, Lirong Cheng, and HY~Fu.
\newblock Edge couplers in silicon photonic integrated circuits: A review.
\newblock {\em Applied Sciences}, 10(4):1538, 2020.

\bibitem{ghione2009semiconductor}
Giovanni Ghione.
\newblock {\em Semiconductor devices for high-speed optoelectronics}.
\newblock Cambridge University Press, 2009.

\bibitem{gopalakrishnan1994performance}
Ganesh~K Gopalakrishnan, William~K Burns, Robert~W McElhanon, Catherine~H
  Bulmer, and Arthur~S Greenblatt.
\newblock Performance and modeling of broadband {LiNbO$_3$} traveling wave
  optical intensity modulators.
\newblock {\em Journal of Lightwave Technology}, 12(10):1807--1819, 1994.

\bibitem{howerton2002broadband}
Marta~M Howerton and William~K Burns.
\newblock {\em Broadband traveling wave modulators in LiNbO$_3$}.
\newblock Cambridge University Press, 2002.

\bibitem{9785841}
Nicola Sambo, Vittorio Curri, Gangxiang Shen, Mattia Cantono, Joao Pedro, and
  Erwan Pincemin.
\newblock Guest editorial: Multi-band optical networks.
\newblock {\em Journal of Lightwave Technology}, 40(11):3360--3363, 2022.

\bibitem{9723008}
Gabriele Di~Rosa, Robert Emmerich, Matheus Sena, Johannes~K. Fischer, Colja
  Schubert, Ronald Freund, and André Richter.
\newblock Characterization, monitoring, and mitigation of the i/q imbalance in
  standard c-band transceivers in multi-band systems.
\newblock {\em Journal of Lightwave Technology}, 40(11):3470--3478, 2022.

\bibitem{9591326}
Robert Emmerich, Matheus Sena, Robert Elschner, Carsten Schmidt-Langhorst,
  Isaac Sackey, Colja Schubert, and Ronald Freund.
\newblock Enabling s-c-l-band systems with standard c-band modulator and
  coherent receiver using coherent system identification and nonlinear
  predistortion.
\newblock {\em Journal of Lightwave Technology}, 40(5):1360--1368, 2022.

\bibitem{honardoost2018high}
Amirmahdi Honardoost, Reza Safian, Ashutosh Rao, and Sasan Fathpour.
\newblock High-speed modeling of ultracompact electrooptic modulators.
\newblock {\em Journal of Lightwave Technology}, 36(24):5893--5902, 2018.

\bibitem{huang2021high}
Xingrui Huang, Yang Liu, Zhiyong Li, Zhongchao Fan, and Weihua Han.
\newblock High-performance and compact integrated photonics platform based on
  silicon rich nitride--lithium niobate on insulator.
\newblock {\em APL Photonics}, 6(11):116102, 2021.

\bibitem{yamada1981dc}
Syoji Yamada and Makoto Minakata.
\newblock {DC} drift phenomena in {LiNbO$_3$} optical waveguide devices.
\newblock {\em Japanese Journal of Applied Physics}, 20(4):733, 1981.

\bibitem{gee1985minimizing}
CM~Gee, GD~Thurmond, H~Blauvelt, and HW~Yen.
\newblock Minimizing dc drift in {LiNbO$_3$} waveguide devices.
\newblock {\em Applied Physics Letters}, 47(3):211--213, 1985.

\bibitem{nagata1993temperature}
Hirotoshi Nagata and Kazumasa Kiuchi.
\newblock Temperature dependence of dc drift of {Ti: LiNbO$_3$} optical
  modulators with sputter deposited {SiO$_2$} buffer layer.
\newblock {\em Journal of Applied Physics}, 73(9):4162--4164, 1993.

\bibitem{salvestrini2011analysis}
Jean~Paul Salvestrini, Laurent Guilbert, Marc Fontana, Mustapha Abarkan, and
  Stephane Gille.
\newblock Analysis and control of the {DC} drift in {LiNbO$_3$} based
  {Mach--Zehnder} modulators.
\newblock {\em Journal of Lightwave Technology}, 29(10):1522--1534, 2011.

\bibitem{nelan2022integrated}
Sean~P Nelan, Andrew Mercante, Shouyuan Shi, Peng Yao, Eliezer Shahid, Benjamin
  Shopp, and Dennis~W Prather.
\newblock Integrated lithium niobate intensity modulator on a silicon handle
  with slow-wave electrodes.
\newblock {\em IEEE Photonics Technology Letters}, 34(18):981--984, 2022.

\bibitem{alam2022net}
Md~Samiul Alam, Essam Berikaa, and David~V Plant.
\newblock Net 350 {Gbps/$\lambda$} {IMDD} transmission enabled by high
  bandwidth thin-film lithium niobate {MZM}.
\newblock {\em IEEE Photonics Technology Letters}, 34(19):1003--1006, 2022.

\end{thebibliography}

\end{document}